\documentstyle[11pt,newpasp,twoside]{article}
\def\lsun{L_{\odot}}
\def\lb{\hbox{$L_{\rm B}$}}

\def\msun{\hbox{$M_\odot$}}

\def\mfecm{\hbox{$M_{\rm Fe}^{\rm ICM}$}}
\def\mfes{\hbox{$M_{\rm Fe}^*$}}
\def\micm{\hbox{$M_{\rm ICM}$}}
\def\mfecm{\hbox{$M_{\rm Fe}^{\rm ICM}$}}

\def\zfe{\hbox{$Z^{\rm Fe}$}}

\def\zfecm{\hbox{$Z^{\rm Fe}_{\rm ICM}$}}
\def\pn{\par\noindent}
\def\spb{\smallskip\pn$\bullet\;$}
\def\edcomment#1{\iffalse\marginpar{\raggedright\sl#1\/}\else\relax\fi}
\catcode`\@=11
\def\gsim{\ifmmode{\mathrel{\mathpalette\@versim>}}
    \else{$\mathrel{\mathpalette\@versim>}$}\fi}
\def\lsim{\ifmmode{\mathrel{\mathpalette\@versim<}}
    \else{$\mathrel{\mathpalette\@versim<}$}\fi}
\def\@versim#1#2{\lower 2.9truept \vbox{\baselineskip 0pt \lineskip 
    0.5truept \ialign{$\m@th#1\hfil##\hfil$\crcr#2\crcr\sim\crcr}}}
\catcode`\@=12
\reversemarginpar

\begin{document}
\title{An Empirical View to the Early Chemical Evolution of Bulge/Disk 
Galaxies}
 \author{Alvio Renzini}
\affil{European Southern Observatory, D-85748, Garching, Germany}
\begin{abstract}
Scaling from the empirical metal yield as measured in clusters of
galaxies, it is inferred that early in the evolution of the Galaxy the
bulge stellar population has produced $\sim 10^9\msun$ of metals, at
least 5 times more than the total metal content of the bulge today. It
is argued that an early galactic wind from the starbursting bulge has
pre-enriched a vast region around it, with these metals being able to
enrich to $\sim 1/10$ solar of order of $5\times 10^{11}\msun$ of
pristine material. From the empirical evidence that bulges come before
disks, it is inferred that the Milky Way disk formed out of this
pre-enriched material, which accounts for the scarcity of metal poor stars in
the solar neighborhood, the so-called
`G-Dwarf Problem'. High redshift observations are now becoming able to 
efficiently explore  the $1.2\lsim z\lsim 3$ region of the universe, when
disk formation and morphological differentiation may have taken place.
\end{abstract}

\section{Introduction}

The chemical evolution of the solar neighborhood, as a prototype of galactic
disks in general, has been an active field of astrophysical research over the 
last four decades, starting from the pioneering works of van den Bergh (1962),
Schmidt (1963), and Pagel \& Patchett (1975). With Talbot \& Arnett (1971) and
then Tinsley (1980), the subject got its elegant physico-mathematical 
formulation, then adopted in most later works. In this traditional approach 
chemical evolution is rigorously treated from first principles, adopting 
specific initial and boundary conditions, along with  heavy element
yields obtained from theoretical stellar model and supernova explosion 
calculations.

In this paper I instead adopt a purely phenomenological approach,
lining up a series of observational facts and then building on them
semi-quantitative inferences. This should be regarded as complementary to the
traditional approach, and may perhaps drive it towards unexplored regions
of the theoretical parameter space. I also adopt a `simple minded' attitude,
assuming -- when appropriate -- that the most straightforward interpretation
of the facts is the right one, and no {\it cosmic conspiracies} are at work
(for example, it is subintended that the IMF is universal).
For conciseness the presentation will be  somewhat schematic.

\section{Learning from Clusters of Galaxies}

Clusters are perhaps the best realization in
nature of the closed-box model of chemical evolution. Within clusters
we find confined in the same place 
all the dark matter, all the baryons, all the galaxies, all the stars and
all the metals, that have participated in the play. Clusters are then good {\it
archives} of their past star formation (SF) and metal production history. 
X-ray observations of clusters provide iron and $\alpha$-element abundances 
in the ICM, along with the ICM mass. Optical observations provide the 
luminosity, mass, and the average metallicity of all the stars now locked
inside galaxies (and even hints to a trace population of intergalactic, 
free-floating stars). Optical and near-IR observations of cluster elliptical 
galaxies have also set tight limits to the formation epoch of the bulk of the
stars contained in them. Cumulatively, these observations have then 
established the 
following facts (see Renzini 1997, 1999a,b and extensive references therein 
for the original sources):
\spb The average iron abundance in the ICM is $\zfecm =(0.3\pm 0.1)Z^{\rm 
     Fe}_\odot$.
\spb ICM {\it iron mass to light ratio} is $\mfecm/\lb = 0.02\pm 0.01\;
       (\msun/\lsun)$,
     where $\mfecm=\zfecm\times\micm$, $\micm$ is the mass of the ICM, and 
     $\lb$ the total $B$-band luminosity of all cluster galaxies.
\spb The average iron abundance of cluster stars is roughly solar. 
\spb The total iron mass to light ratio is $(\mfes + \mfecm)/\lb=0.03\pm 0.01\;
       (\msun/\lsun)$, only weakly dependent on the Hubble constant.
\spb The global elemental ratio $[\alpha /$Fe] in the ICM+stars is roughly 
     solar, hence the total cluster {\it metal mass to light ratio} is
     $M_{\rm Z}/\lb=0.3\pm 0.1\; (\msun/\lsun)$, since in solar proportions
     $Z\simeq 10\times\zfe$. This is a fully {\it empirical} estimate of the 
     metal yield of a now old stellar population.
\spb Most metals are out of galaxies: $\mfecm/\mfes\simeq 1.6\, h^{-3/2}$,
     i.e. there is $\sim 2.5$ times more iron in the ICM than there is
     iron locked into stars,
     for $H_\circ=75$, or $\sim 4.5$ times more for  $H_\circ=50$.
\spb Massive starbursts promote major, metal-enriched galactic winds (Heckman 
     et al. 2000).
\spb Most stars in galaxy clusters belong to galactic spheroids (ellipticals 
     and bulges), and formed in massive starbursts at $z\gsim 3$.
\par\medskip
Note that in $M_{\rm Z}/\lb$ both quantities are measured {\it now}; however,
the metal mass $M_{\rm Z}$ was produced and released at very early times by the
stellar population that after aging for $\sim 13$ Gyr has faded to the 
luminosity $\lb$.
In the adopted `simple-minded' approach, from these empirical facts the 
following inferences can be drawn:
\smallskip\pn
$\star$ Most metals in clusters (now partly in the ICM, partly in stars) were
        produced at $z\gsim 3$.
\smallskip\pn
$\star$ ICM metals were ejected by starburst driven galactic winds at the
production time, i.e. at $z\gsim 3$.

\section{Field vs Clusters}

To which extent are clusters {\it fair samples} of the universe as a whole?
They are certainly the highest density peaks in the distribution of matter,
both dark and shining, and one may expect SF and chemical 
evolution to have proceed differently in clusters compared to the low-density
{\it field}. However, there are also striking similarities, including the
following ones:
\spb Most stars are now in galactic spheroids, in the field as in the clusters
     (e.g. up to $\sim 75\%$ according to Fukugita, Hogan, \& Peebles 1998).
\spb The Fraction of baryons now locked into stars is nearly independent of 
     the environment, being $\sim 10\%$ in both clusters and field (e.g.
     Renzini 1997; Fukugita et al. 1998). The SF histories may well have been 
     different, confined at early times in clusters, more protracted in the 
     field, but the $z=0$ endproducts appear to be very similar, with nearly 
     the same efficiency of baryon to star conversion.
\spb Field ellipticals and S0's are very similar to cluster ellipticals and
     S0's, with the luminosity-weighed age of their stellar populations being
     at most $\sim 1$ Gyr less than that of the cluster galaxies (Bernardi
     et al. 1998). 
\spb The stellar populations of (large) bulges are very similar to 
     ellipticals. The majority of them appear to follow the same Mg$_2-\sigma$
     relation of ellipticals, while a minority of outliers likely had a
     significant episode of star SF at later times (Jablonka, Martin, 
     \& Arimoto 1996).
\spb The stellar population of the Galactic bulge is nearly as old as Galactic
     globular clusters in halo, or 13-15 Gyr (Ortolani et al. 1995).
\pn
\medskip
Again, from this second series of facts more `simple-minded' inferences can
be drawn:
\smallskip\pn 
$\star$ Having the SF proceeded to the same $\sim 10\%$
baryon to stars conversion in the general field and in clusters, the
global metallicity of the $z=0$ universe has to be nearly the same as that we 
can actually see in clusters, i.e. $\sim 1/3$ solar.
\smallskip\pn 
$\star$ The bulk of stars in galactic spheroids -- in clusters as well as in 
the field -- formed at $z\gsim 3$.  
\smallskip\pn 
$\star$ Since some 50 to 75\% of all stars are now in spheroid, and the bulk 
of stars in spheroids formed at $z\gsim 3$, the inference 
is that at least $\sim 1/3$ of all stars formed at $z\gsim 3$. This `fossil 
evidence' argument agrees with the direct determination of the SF history at 
high redshift (Steidel et al. 1999).
\smallskip\pn 
$\star$ Therefore, the metallicity of the $z=3$ universe was $\sim 1/3$ of its
 present value, i.e. $\sim 1/3\, \cdot \, 1/3\simeq 1/10$ solar, a {\it
prompt initial enrichment of the universe} (Renzini 1999a).

\section{Bulges vs Disks}

When did the morphological differentiation of galaxies take place? When did 
the 
disks of the present-day spirals started to be assembled? Here are some facts:
\spb We see no disk galaxies at $z\gsim 3$. Lyman break galaxies appear to be 
much
smaller, compact objects, most likely the progenitors of today's objects that
the fossil evidence date having formed at $z\gsim 3$, i.e. galactic spheroids
(e.g. Giavalisco, Steidel, \& Macchetto 1996).
\spb Spirals are well in place by $z=1$, along with passively evolving 
ellipticals (e.g. Abraham \& van den Bergh 2001).
\medskip\pn
Hence:
\smallskip\pn 
$\star$ Bulges come first, formed in starbursts, and disks are slowly 
added later, if the environment is quiet enough to allow for their
 formation and survival.
\smallskip\pn 
$\star$ The morphological differentiation of galaxies (the emergence of the
Hubble sequence) took place between $z\sim 3$ and $z\sim 1$, i.e. in the 
so far poorly explored {\it desert} range $1\lsim z\lsim 3$.

\section{The Early Chemical Evolution of the Milky Way}

In the K band the Galactic bulge and disk contribute respectively
$\sim 1.2\times 10^{10}$ and $\sim 5.5\times 10^{10}\; L_{\rm
K,\odot}$ (Kent, Dame, \& Fazio 1991), and in the $B$ band the bulge
luminosity is $L^{\rm BULGE}_{\rm B}\simeq 6\times 10^9 L_{\rm B,
\odot}$.

>From the cluster empirical yield it follows that the Galactic bulge has 
produced $M_{\rm Z}\simeq 0.3L^{\rm BULGE}_{\rm B}=0.3\times 6\times 
10^9\simeq 2
\times 10^9\msun$ of metals. Where are all these metals? Two billion solar
masses of metals should not be easy to hide, yet ...
The stellar mass of the bulge follows from its $K$-band mass to light ratio,
$M_*^{\rm BULGE}/L_{\rm K}=1$ (Kent 1992), and its luminosity,
hence $M_*^{\rm BULGE}
\simeq 10^{10}\msun$. Its average metallicity is about solar(McWilliam \& 
Rich 1994), i.e. $Z=0.02$, and therefore the bulge stars all together contain
$\sim 2\times 10^8\msun$ of metals. Only $\sim 1/10$ of the metals 
produced when the bulge was actively star forming some 13 Gyr ago are still
in the bulge! This implies that $\sim 90\%$, or $\sim 1.8\times 10^9\msun$ 
were ejected into the surrounding space by an early wind.

A word of caution is in order. The estimated the metal yield (Section 2)
follows from adopting $\zfecm=0.3$ solar. More recent estimates prefer
$\zfecm=0.2$ solar (De Grandi \& Molendi 2001), the total metal production 
by the bulge reduces to $\sim 1.3\times 10^9\msun$, of which $\sim 10^9\msun$
had to be ejected. The bottom line is that at least $\sim 5$ times more 
metals were ejected, than retained in the bulge.

At the time of bulge formation, such $\sim 10^9\msun$ of metals run
into largely pristine ($Z=0$) material, experienced R-T instabilities
leading to chaotic mixing, and establishing a distribution of
metallicities in a largely inhomogeneous IGM surrounding the young
bulge. For example, this enormous amount of metals is able to bring to
a metallicity 1/10 solar (i.e. $Z=0.002$) about $5\times 10^{11}\msun$
of pristine material, several times the mass of the yet to be formed
Galactic disk.

\subsection{Three Phases in the Milky Way Build Up} 

If we look to the formation of the Milky Way galaxy from a purely empirical
point of view, we have an old, now passively evolving bulge, and a younger
disk, still forming stars even if -- likely -- at a reduced rate compared to 
a more active past. Hence, we can distinguish three main phases in the
formation and evolution of our own Galaxy.
\medskip\pn
{\bf Phase 1: Bulge Formation}, some 13 Gyr ago (``at $z\sim 3$''). The 
relatively fast ($\lsim 1$ Gyr) assembly of the bulge from smaller subunits
promotes a massive starburst (SFR=10-100 $\msun/$yr), which drives a metal
rich wind and $\sim 10^9\msun$ of metals are ejected into the surrounding
medium.
\medskip\pn {\bf Phase 2: The Intermission}, lasting a poorly
constrained lapse of time (some Gyr?, until $z\sim 2.5$?), during which
the bulge settles into its passive evolution, the ejecta partially mix
with the surrounding medium, and a mass some 5-10 times larger than
the present mass of the Galaxy is inhomogeneously contaminated, with
its average metallicity being raised to $\sim 1/10$ solar. Heating of
this metal enriched environment (MEE) by the early bulge wind, may
prevent from a while further growth of the stellar mass of the
protogalaxy.  
\medskip\pn{\bf Phase 3: Disk Formation}, from $z\sim 2.5$ to
$z\sim 1$ (?). Around the aging bulge, the MEE is cooling, infall
starts of $Z\simeq 1/10$ solar material from the MEE, and the Galactic
disk begins to form and grow.
 
\section{Conclusions}

These semi-quantitative arguments may help settling on a final solution
for an old problem. From the earliest times mentioned in the Introduction,
the existence of a ``G-Dwarf Problem'' was soon recognized: the solar
neighborhood contains far too few metal poor stars ($z<1/10$ solar) compared
to the prediction of a simple, closed box model of chemical evolution
(see Pagel 2001, for a recent discussion of the problem). Traditional 
solutions of the G-Dwarf problem have considered three options (or some
combination of them): an initial {\it top heavy} IMF, a {\it prompt
initial enrichment} (PIE) model, and infall, i.e. the gradual assembly
of the disk as opposed to a zero metallicity disk fully assembled from the 
beginning. The above considerations clearly indicate that PIE is inevitable,
once one accepts that the bulge comes first and the disk later. Note that
infall models have often assumed $Z=0$ for the infalling material.
Some gradual
growth of the disk is also inevitable, being the alternative assumption of 
instantly assembled disk quite unrealistic. So, a combination of PIE and
infall emerges as the natural solution of the old problem. Worth
mentioning here is that the idea of the bulge pre-enriching the disk
is certainly not new (see e.g. K\"oppen \& Arimoto 1990).  

This sketchy cartoon of galaxy formation is hard to put in more
quantitative terms, purely from first principles. Models would have 
difficulties to
predict the extension of the MEE, its distribution of metallicities,
the duration of the intermission, the evolution of the infall rate
and the disk build up. All these phenomena involve highly non-linear, 
weather-like hydrodynamics with little predictive power. 
However, we are living in a special time for 
astronomy, and where theory gets stuck observations can help. We can  
actually start to {\it see} directly what theory is not able to predict.

With the current generation of large ground based telescopes, together with 
the X-ray, optical, and infrared facilities now or soon in space, we have
for the first time the possibility to empirically map the growth of
galaxies, to follow back in time the disappearance of the Hubble sequence,
and its emergence, forwards from high to low redshifts.

\end{document}